\documentclass{iau} 
\usepackage{aas_macros}
\usepackage{hyperref}
\usepackage{url}
\usepackage{enumitem}
\usepackage{mathtools}
\usepackage{graphicx}
\usepackage{amssymb}
\usepackage{amsmath}
\usepackage{ulem}
\usepackage{epstopdf}
\usepackage{color}
\def\teq#1{$\, #1\,$}      
\def\erg{\varepsilon} 

\usepackage{graphicx}

\title[Magnetar Hard Spectral Tails] 
{Hard Tails in Magnetars}

\author[Wadiasingh et al.]   
{\vspace{-2.8mm} Zorawar Wadiasingh$^1$, Matthew G. Baring$^2$, Peter L. Gonthier$^3$, Alice K. Harding$^4$ \vspace{-0mm}}

\affiliation{$^1$Centre For Space Research, North-West University, Potchefstroom, South Africa \\
$^2$Rice University, Houston, Texas, USA \\
$^3$Hope College, Holland, Michigan, USA\\
$^4$NASA Goddard Space Flight Center, Greenbelt, Maryland, USA 
\vspace{-0mm}}

\pubyear{2017}
\volume{337}  
\jname{Pulsar Astrophysics -­ The Next 50 Years}
\editors{P. Weltevrede, B.B.P. Perera, L. Levin Preston \& S. Sanidas, eds.}

\begin{document}

\maketitle
\vspace{-4mm}
\begin{abstract}
Pulsed non-thermal quiescent emission between 10 keV and around 150 keV has been observed in $\sim10$ magnetars. For inner magnetospheric models of such hard X-ray signals, resonant Compton upscattering of soft thermal photons from the neutron star surface is the most efficient radiative process. We present angle-dependent hard X-ray upscattering model spectra for uncooled monoenergetic relativistic electrons. The spectral cut-off energies are critically dependent on the observer viewing angles and electron Lorentz factor. We find that electrons with energies less than around 15 MeV will emit most of their radiation below 250 keV, consistent with the observed turnovers in magnetar hard X-ray tails. Moreover, electrons of higher energy still emit most of the radiation below around 1 MeV, except for quasi-equatorial emission locales for select pulses phases. Our spectral computations use new state-of-the-art, spin-dependent formalism for the QED Compton scattering cross section in strong magnetic fields. 
\vspace{-2mm}

\vspace{-3mm}
\end{abstract}

\firstsection 
\vspace{-5mm}
\section{Background}
\vspace{-1mm}

\fontdimen2\font=2.5pt

Magnetars are a group of slowly-rotating $P \sim 2- 12$ s young neutron stars with exceptionally high inferred surface magnetic fields. They are also distinguished by their super-Eddington bursting activity and persistent emission exceeding spin-down power by orders of magnitude. The latter is a key indicator for dissipation of magnetic fields as the energy source. Magnetar giant flares are among the most powerful high-energy events ever observed at Earth. For instance, the 2004 giant flare of SGR 1806-20, which exceeded $10^{47}$ erg s$^{-1}$ briefly, caused measurable changes in the Earth's daytime ionosphere (\cite{2007GeoRL..34.8103I}). The surface dipole fields of magnetars are usually deduced from X-ray timing of spin period and period derivatives via the standard vacuum dipole formula $B_p = 6.4 \times 10^{19} (P \dot{P})^{1/2}$. Magnetar fields may exceed the quantum critical or Schwinger field $B_{cr}$, where the cyclotron energy of an electron is equal to its rest mass energy, $B_{\rm cr} = m^2 c^3/( e \hbar) \approx 4.413 \times 10^{13}$~G. Exotic quantum electrodynamic (QED) processes can play a major role in spectral formation for fields near $B_{\rm cr}$ (see \cite{2006RPPh...69.2631H}). 

 Some high-field rotation-powered pulsars with $P\lesssim 1$ s, e.g. PSR J1119-6127 and PSR J1846-0258, also exhibit sporadic magnetar-like behavior. The magnetic dipole spin-down formula is not solely determinative of the field value since it may miss strong multipolar components as exemplified by the ``low-field" magnetar SGR 0418+5729, which has a spin-down torque dipole field of ``only" $6 \times 10^{12}$ G yet exhibits magnetar activity (\cite{2013ApJ...770...65R}).  Yet, it presents a proton cyclotron absorption features indicative of magnetar-like surface fields (\cite{2013Natur.500..312T}). Currently, around 30 magnetars are known\footnote{The McGill magnetar catalog: \\ \url{http://www.physics.mcgill.ca/~pulsar/magnetar/main.html}}. For recent reviews, see \cite{2008A&ARv..15..225M}, \cite{2015RPPh...78k6901T} and \cite{2017ARA&A..55..261K}. 

In the X-rays, magnetars present two distinct persistent emission components: soft quasi-thermal emission up to around $10$ keV, and very flat nonthermal hard X-ray tails $dN/dE \propto E^{-\Gamma}$ with $\Gamma_h \sim 0.5-1$ extending to beyond 200 keV. Both persistent components may exhibit luminosities of $10^{33}-10^{36}$ erg s$^{-1}$ well in excess of the dipole spin-down value. The soft X-ray emission is thought to primarily originate from the hot surface of temperature $T \sim 10^{7}$ K, supplemented by resonant cyclotron Comptonization by mildly relativistic electrons (\cite{2006MNRAS.368..690L}). Hard tails have been detected in about 10 magnetars by RXTE, INTEGRAL, Suzaku, Fermi-GBM and NuSTAR. Archival data from the COMPTEL instrument on the CGRO constrains the extension of these power laws to $\sim 500$ keV (e.g. \cite{2008A&A...489..245D}) where it must peak in a $\nu F_\nu$ representation. Therefore the hard tails may dominate the emission energetics. Such spectral turnovers are reinforced by upper limits in \textit{Fermi} LAT data above 100 MeV for several magnetars (\cite{2010ApJ...725L..73A}, \cite{2017ApJ...835...30L}). 

There is rich complexity in the persistent X-ray emission pulse profiles of many magnetars, particularly in epochs following burst activity (e.g. \cite{2015ApJ...807...93A}). However, the macrostructure of profiles exhibit broad and energy-dependent single or double peaks, with total pulsed fraction and hardness of emission increasing with energy. Such broad pulses are in sharp contrast with narrow peaks observed in rotation-powered radio and $\gamma$-ray pulsars, and is suggestive of magnetar emission locales occurring in the inner {\textit{closed}} magnetosphere, not far from the neutron star surface. 

For the hard X-ray domain, relativistic electrons are essential in the closed magnetosphere. The electrons are presumed to be accelerated along closed field lines, by static electric potentials, or dynamic ones associated with large scale currents and twists in the magnetic field (e.g. \cite{2005ApJ...634..565T}, \cite{2007ApJ...657..967B}).  Understanding the activation of the magnetosphere is a difficult problem that likely entails magnetic pair cascades coupling to acceleration regions in a nonlinear manner. Our upcoming extensive work \cite{wadiasinghsubitted} (hereafter WBGH17) focuses on the first-principles radiation modeling, presuming a source of relativistic electrons. This work forms an integral element of future models of self-consistent kinetic particle dynamics. Regardless of the radiative process, the hard tails require particle densities far in excess of the magnetar's Goldreich-Julian values. Spectropolarimetric radiation modeling therefore enables key diagnostics on the emission locales.

\vspace{-7.2mm}
\section{Resonant Inverse Magnetic Compton Scattering}
\vspace{-1mm}

In strong magnetic fields, electron states are described by discrete Landau levels. The cyclotron lifetime of electrons is very short, $\sim 10^{-19}$ s, quickly radiating away electron momentum perpendicular to the local field. Consequently, most electrons occupy the ground Landau state and are constrained to move along field lines. The intense surface thermal photon bath enables resonant inverse Compton scattering by relativistic electrons, which is effectively a first-order QED process of cyclotron absorption followed by emission from the intermediate Landau state. We, and others, have shown that resonant inverse Compton scattering of thermal X-rays is the dominant radiative mechanism for ultrarelativistic electrons in magnetar magnetospheres (\cite{2007Ap&SS.308..109B}, \cite{2007ApJ...660..615F}, \cite{2011ApJ...733...61B}). Depending on the magnetospheric locale and electron directionality, cooling lengths may be short, $\ll 10^6$ cm. This is particularly pertinent in the hard tail context, since scattered photon energies may approach $ \erg_f \sim m_e c^2 \gamma_e B/B_{\rm cr}$ for electron Lorentz factor $\gamma_e$.

\begin{figure}[t]
\centering
 \includegraphics[width=0.9\textwidth]{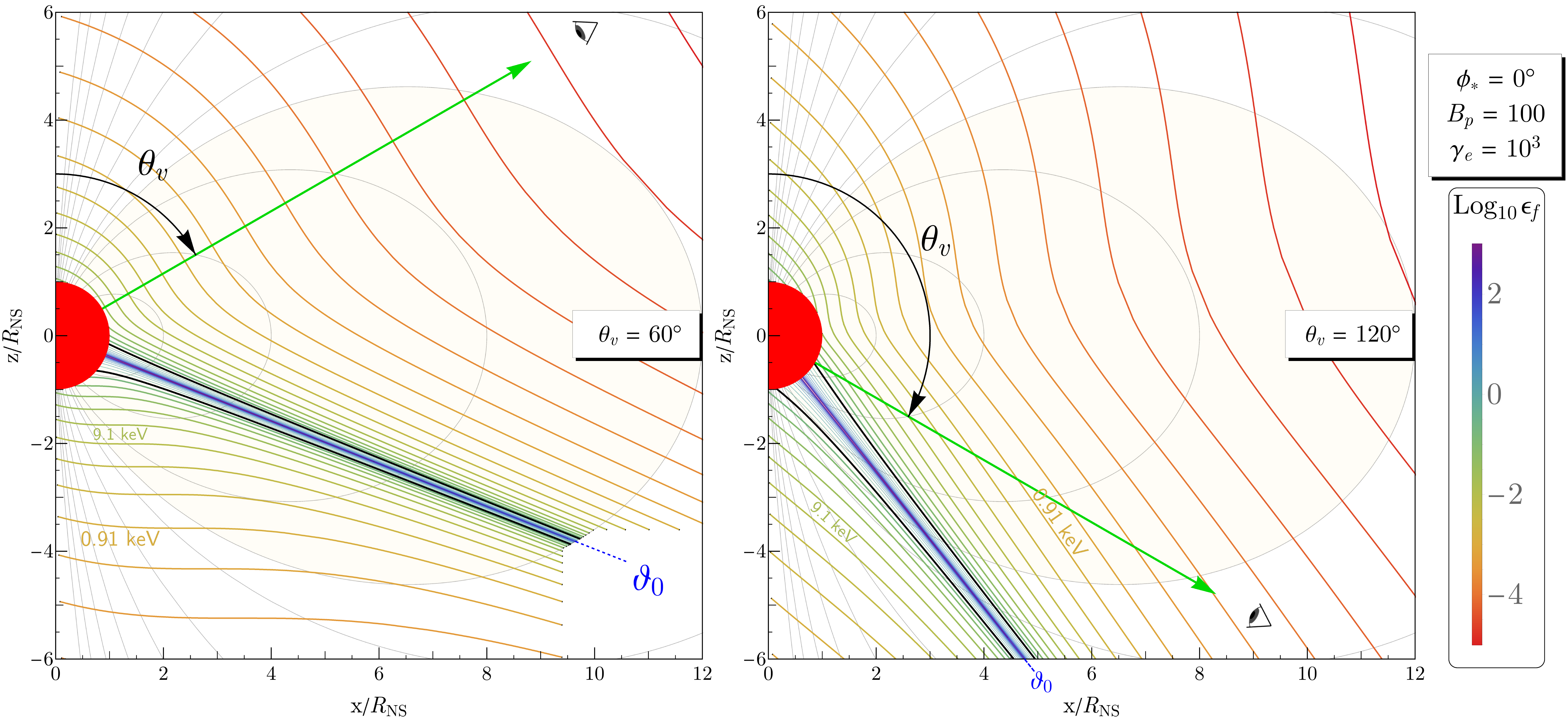} 
 \caption{Resonance interaction loci along meridional field loops for observer viewing angles \teq{\theta_v = 60^{\circ}} and \teq{\theta_v = 120^{\circ}}. As the magnetic field drops with increasing altitude, the interaction points converge towards a radial dotted blue line with the colatitude \teq{\vartheta_0}. }
    \vspace{-3mm}
   \label{fig1}
\end{figure}

Resonant Compton upscattering emission is necessarily anisotropic, due to Doppler boosting mostly into a narrow Lorentz cone that is a key ingredient in sampling the resonant character. In WBGH17, we examine a first-principles model of directed anisotropic emission to capture the essential character of the emission, where electrons of fixed Lorentz factor are constrained along field loops in a dipole magnetosphere. The inversion of the kinematics for resonant interactions yields a one-to-one correspondence between the final scattering angle and observer-frame scattered energy, yielding a relatively small spatial locale where the highest energy emission is beamed toward an instantaneous observer line-of-sight that varies with rotation phase. 

In Fig. \ref{fig1}, we depict such resonant interactions the meridional plane (the plane formed by the observer vector and dipole axis) of a dipole magnetic field for two observer viewing directions where electrons of $\gamma_e = 10^{3}$ are presumed to move from the south to north magnetic pole. Here $B_p$ is the polar field in units of $B_{cr}$. The hardest emission is beamed near field tangents directed at the observer, forming an observer-dependent ``separatrix" at $\vartheta_0$.  Contours of constant $\Psi \equiv B_p/[2\erg_f \gamma_e]$ are depicted, with a range \teq{10^{-4} < \Psi < 4 \times 10^3},  being color-coded according to the  value of the dimensionless final photon energy \teq{\erg_f}, as indicated in the legend on the right;  the black loci constitute the value of  \teq{\erg_f=10^{-0.5}\approx 160}keV (i.e., \teq{\Psi \approx 0.16}). For twisted field geometries that depart from a dipole, such resonant zones will be different. However the dipole construction is still likely fair at higher-altitude equatorial zones pertinent for hard tails. As the magnetar spins with angle $\alpha$ between spin and magnetic axes, different equatorial locales and activated regions are sampled, effectively generating pulses (see WBGH17).

\begin{figure}[t]
\centering
 \includegraphics[width=0.9\textwidth]{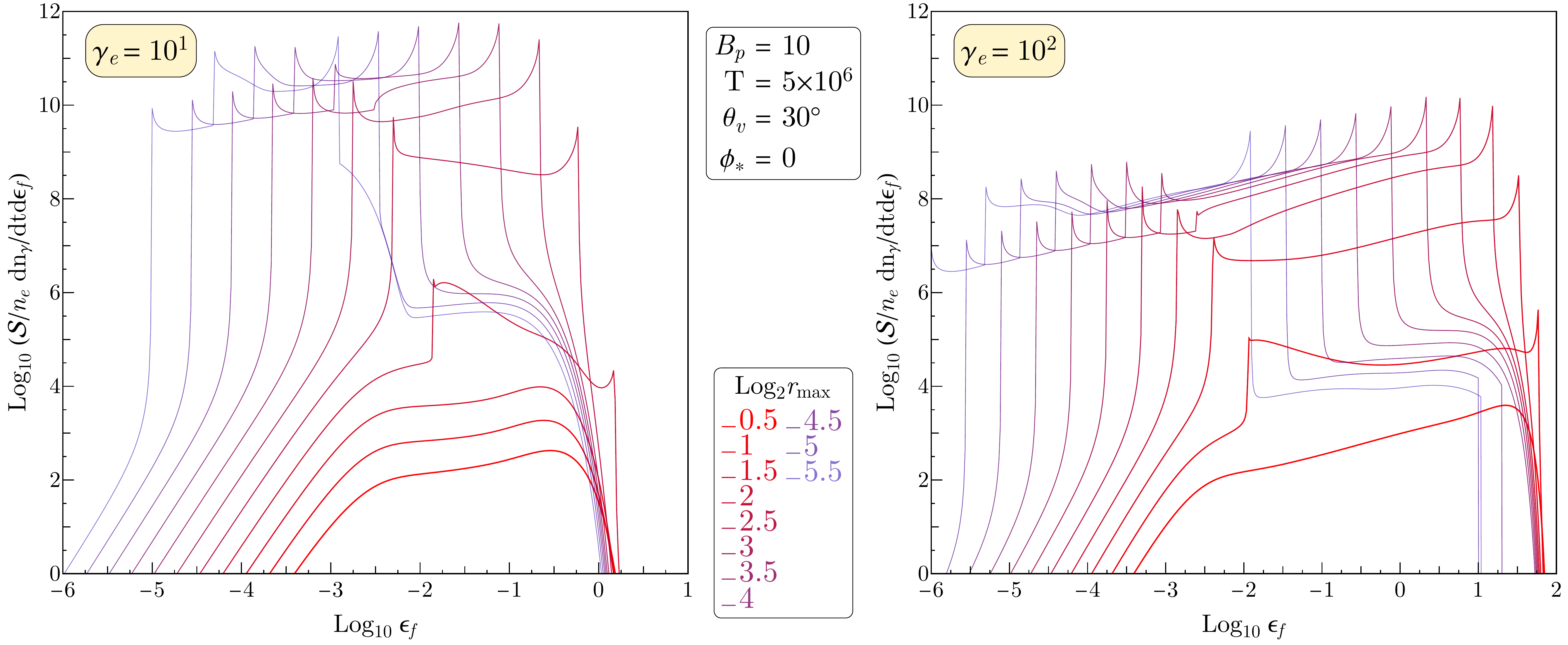} 
 \caption{Upscattering spectra for meridional field loops depicting several choices of the maximum loop altitude parameter 
\teq{r_{\rm max} = \{2^{0.5},...,2^{5.5}\}}, for surface temperature $T = 5\times 10^6$ K. The two panels are for two different 
Lorentz factors, and spectra are realized for a viewing angle of 
\teq{\theta_v = 30^\circ}. Clear variation in the cut-off energies and normalization are evident, 
as well as a transition from weakly-resonant to fully-resonant from low to 
moderate and higher altitudes.}
    \vspace{-2mm}
   \label{fig2}
\end{figure}

Kinematic Lorentz transformations along with resonance sampling criteria, combined with the appropriate spin-dependent Sokolov and Ternov QED cross section (\cite{2014PhRvD..90d3014G}) and inverse lifetime to ``cap" the resonance (\cite{2005ApJ...630..430B}) regulate photon spectra. Collision integrals, effectually formulated in terms of differential photon production rate (or scattering rate) $ dn_\gamma/ (d t d \erg_f d \mu_f) \sim c \int \sigma n_\gamma $ are integrated over initial and final scattering angles and electron paths. A small sampling of such differential spectral computations in WBGH17 is depicted in Fig. \ref{fig2} for single meridional field loops at various altitudes, coupling to similar resonant locales as Fig.~\ref{fig1}. When resonant interactions are effectively sampled, i.e. between the cusps, differential spectra attain $dn_\gamma/ (d t d \erg_f) \sim \erg_f^{1/2}$, flatter than what is observed. The higher $\gamma_e = 10^{2}$ violates cut-off bounds for higher Lorentz factors for some altitudes. This then suggests $\gamma_e \lesssim 30$, generally consistent with cooling rate calculation expectations. For some low altitudes (high local $B$), resonances are not effectively sampled, suppressed by the lower numbers of photons in the Wien tail of the surface thermal Planckian. Yet at higher altitudes, spectra cut-off at low energies, the regime where the surface thermal X-rays dominate. The superposition of these curves, as well as off-meridional spectra depicted in WBGH17, give a strong indication of spectra that might result from toroidal volumes to match observations. Moreover, as the magnetic Thomson results of \cite{2013ApJ...762...13B} suggest, self-consistent cooling dynamics of electrons is an essential next step.

In addition to spectra at instantaneous viewing angles, we compute pulse profiles in WBGH17 presented in terms of sky maps in pulsar $\zeta$. The pulsations are generally broad and single-peaked, yet exhibit symmetric double-peak structure near $\alpha \sim \zeta$. These are viewing directions that sample the magnetic pole and equatorial locales of resonant scattering. The double-peak separation generally decreases with higher altitudes or upscattered photon energies. Application to 1E 1841-045 in WBGH17 suggests $\alpha \lesssim 20^\circ$, generally consistent with $\alpha \sim 15^\circ$ inferred by \cite{2014ApJ...786L...1H}.

\vspace{-7.2mm}
\section{Future}
\vspace{-1mm}

Our WBGH17 radiation modeling sets the stage for future kinetic simulations of electron dynamics and photon radiative transport in magnetars. This will allow for useful constraints on magnetar $\alpha$. QED magnetic pair production and photon splitting are also expected to attentuate emergent spectra in a phase- and polarization-dependent manner. Spectropolarimetric observations with future space-borne instruments will usher in the era of astrophysical tests of high-field QED in the next decade\footnote{See, for example, AMEGO: \url{https://asd.gsfc.nasa.gov/amego/index.html}}.



\vspace{-7.2mm}
\begin{thebibliography}{}  
\scriptsize
\setlength{\itemsep}{-0.2mm}
\vspace{-1mm}

\bibitem[Abdo et al. 2010]{2010ApJ...725L..73A} Abdo, A.~A., Ackermann, 
M., Ajello, M., et al.\ 2010, \apjl, 725, L73 

\bibitem[An et al. 2015]{2015ApJ...807...93A} An, H., Archibald, R.~F., Hasco{\"e}t, R., et al.\ 2015, \apj, 807, 93 


\bibitem[Baring 
\& Harding 2007]{2007Ap&SS.308..109B} Baring, M.~G., \& Harding, A.~K.\ 2007, \apss, 308, 109 


\bibitem[Baring et al. 2005]{2005ApJ...630..430B} Baring, M.~G., Gonthier, P.~L., \& Harding, A.~K.\ 2005, \apj, 630, 430 

\bibitem[Baring et al. 2011]{2011ApJ...733...61B} Baring, M.~G., Wadiasingh, Z., \& Gonthier, P.~L.\ 2011, \apj, 733, 61 

\bibitem[Beloborodov (2013)]{2013ApJ...762...13B} Beloborodov, A.~M.\ 2013, \apj, 762, 13 

\bibitem[Beloborodov \& Thompson 2007]{2007ApJ...657..967B} Beloborodov, A.~M., \& Thompson, C.\ 2007, \apj, 657, 967 

\bibitem[Fern{\'a}ndez \& Thompson 2007]{2007ApJ...660..615F} Fern{\'a}ndez, R., \& Thompson, C.\ 2007, \apj, 660, 615 

\bibitem[Gonthier et al. 2014]{2014PhRvD..90d3014G} Gonthier, P.~L., Baring, M.~G., Eiles, M.~T., et al.\ 2014, \prd, 90, 043014 

\bibitem[Harding \& Lai 2006]{2006RPPh...69.2631H} Harding, A.~K., \& Lai, D.\ 2006, Reports on Progress in Physics, 69, 2631 

\bibitem[den Hartog et 
al. 2008]{2008A&A...489..245D} den Hartog, P.~R., Kuiper, L., Hermsen, W., et al.\ 2008a, \aap, 489, 245 

\bibitem[Hasco{\"e}t et al. (2014)]{2014ApJ...786L...1H} Hasco{\"e}t, R., Beloborodov, A.~M., \& den Hartog, P.~R.\ 2014, \apjl, 786, L1 



\bibitem[Inan et al. 2007]{2007GeoRL..34.8103I} Inan, U.~S., Lehtinen, N.~G., Moore, R.~C., et al.\ 2007, \grl, 34, L08103 

\bibitem[Kaspi \& Beloborodov (2017)]{2017ARA&A..55..261K} Kaspi, V.~M., \& Beloborodov, A.~M.\ 2017, \araa, 55, 261 

\bibitem[Li et al. 2017]{2017ApJ...835...30L} Li, J., Rea, N., Torres, D.~F., \& de O{\~n}a-Wilhelmi, E.\ 2017, \apj, 835, 30 

\bibitem[Lyutikov \& Gavriil 2006]{2006MNRAS.368..690L} Lyutikov, M., \& Gavriil, F.~P.\ 2006, \mnras, 368, 690 

\bibitem[Mereghetti (2008)]{2008A&ARv..15..225M} Mereghetti, S.\ 2008, \aapr, 15, 225 

\bibitem[Rea et al. 2013]{2013ApJ...770...65R} Rea, N., Israel, G.~L., 
Pons, J.~A., et al.\ 2013, \apj, 770, 65 

\bibitem[Thompson \& Beloborodov 2005]{2005ApJ...634..565T} Thompson, C., \& Beloborodov, A.~M.\ 2005, \apj, 634, 565 

\bibitem[Tiengo et al. 2013]{2013Natur.500..312T} Tiengo, A., Esposito, 
P., Mereghetti, S., et al.\ 2013, \nat, 500, 312 

\bibitem[Turolla et al. (2015)]{2015RPPh...78k6901T} Turolla, R., Zane, S., \& Watts, A.~L.\ 2015, Reports on Progress in Physics, 78, 116901 

\bibitem[Wadiasingh et al. (2017)]{wadiasinghsubitted}  Wadiasingh, Z., Baring, M.~G., Gonthier, P.~L., \& Harding, A.~K.\ 2017, submitted to \apj, [WGBH17]

\end{thebibliography}
\end{document}